\documentclass[prd,preprint,superscriptaddress,amsmath,amssymb,nofootinbib]{revtex4}
\usepackage{graphicx}
\usepackage{dcolumn}
\usepackage{bm}
\usepackage{amssymb}
\usepackage{amsmath}
\usepackage{epsfig}    
\usepackage{color}
\usepackage{slashed}
\usepackage{hhline}

\def\be{\begin{equation}}
\def\ee{\end{equation}}
\newcommand{\bea}{\begin{eqnarray}}
\newcommand{\eea}{\end{eqnarray}}

\begin{document}

\title{Constraining vector dark matter and dark photon with degenerate mass in a hidden local $SU(2)$ model}

\author{Takaaki Nomura}
\email{nomura@scu.edu.cn}
\affiliation{College of Physics, Sichuan University, Chengdu 610065, China}

\author{Xinran Xu}
\email{xinran\_xu@stu.scu.edu.cn}
\affiliation{College of Physics, Sichuan University, Chengdu 610065, China}

\date{\today}

\begin{abstract}
We discuss degenerate vector dark matter and dark photon that are induced from hidden $SU(2)_H$ gauge sector where it is spontaneously broken by vacuum expectation value of $SU(2)_H$ doublet.
Kinetic mixing between $SU(2)_H$ and $U(1)_Y$ gauge fields can be generated by introducing dimension six operator realizing dark photon interactions.
In estimating relic density we focus on the process in which dark matter annihilate into dark photons, and search for the region of dark matter mass and gauge coupling realizing observed relic density.
We then discuss constraints from dark photon physics, thermalization of dark sector and direct detection of dark matter.
It is then found that constraints from direct detection experiments give us the strongest upper limits on the dark photon interactions. 

 \end{abstract}
\maketitle

\section{Introduction}

Understanding of dark matter (DM) nature is one of the biggest challenge in physics.
The existence of DM definitely requires physics beyond the standard model (SM) where we call new physics sector containing DM as dark sector.  
In fact various candidates of DM are considered from new physics models.
One promising candidate of DM is a massive stable particle that weakly interacts with SM sector as its relic density can be explained via thermal production.
Such a DM has been searched for by direct detection, indirect detection and collider experiments.
However we do not have any clear evidence of DM yet and nature of dark sector is an open question.

One attractive scenario is to consider dark sector as hidden gauge sector as the SM is also described by gauge symmetry.
We consider hidden gauge symmetry where SM particles are neutral but fields in dark sector including DM are charged under it.
In this kind of scenario we can explain stability of DM by a remnant symmetry after spontaneous symmetry breaking of hidden gauge symmetry~\cite{Krauss:1988zc, Ko:2018qxz}.
In particular it is interesting to consider minimal non-Abelian group $SU(2)_H$ as hidden gauge symmetry.
In the minimal scenario~\cite{Hambye:2008bq},  $SU(2)_H$ is broken by vacuum expectation value (VEV) of $SU(2)_H$ doublet scalar $\Phi$ and dark gauge bosons are DM candidate 
where the stability of them is guaranteed by remaining custodial symmetry~\footnote{There are various approaches applying a hidden $SU(2)$ gauge symmetry in literatures; A remaining $Z_{2,3,4}$ symmetry with a quadruplet(quintet) scalar in ref.~\cite{Chiang:2013kqa, Chen:2015nea,Chen:2015dea,Chen:2015cqa,Ko:2020qlt,Nomura:2020zlm,Chen:2017tva}, $Z_2 \times Z'_2$ symmetry~\cite{Gross:2015cwa}, a custodial symmetry in refs.~\cite{Boehm:2014bia, Hambye:2008bq,Baouche:2021wwa}, an unbroken $U(1)$ from $SU(2)$ in refs.~\cite{Baek:2013dwa,Khoze:2014woa,Daido:2019tbm}, a model adding hidden $U(1)_h$~\cite{Davoudiasl:2013jma, Choi:2019zeb}, other DM scenarios~\cite{Zhang:2009dd, Barman:2017yzr,Barman:2018esi,Barman:2019lvm,Barman:2020ifq}, a model with classical scale invariance~\cite{Karam:2015jta}, Baryogengesis~\cite{Hall:2019ank}, connection to neutrino mass generation~\cite{Nomura:2021aep}, electroweak phase transition~\cite{Ghosh:2020ipy} and some general discussion~\cite{Yuan:2024tfs}.}.
Interestingly, we can have kinetic mixing between $SU(2)_H$ and $U(1)_Y$ gauge fields if we introduce dimension 6 operator of $(\Phi^\dagger \sigma^a \Phi)X^a_{\mu \nu} B^{\mu \nu}$ where  $X_{\mu \nu}$ and $B_{\mu \nu}$ are the gauge field strength of $SU(2)_H$ and $U(1)_Y$ respectively and $\sigma^a$ is the Pauli matrix on $SU(2)_H$ representation space~\cite{Zhou:2022pom}~\footnote{A mixing between $SU(2)_L$ and hidden $U(1)$ is also discussed in ref.~\cite{Arguelles:2016ney}}; Dimension 5 operator is also possible introducing $SU(2)_H$ triplet scalar~\cite{Zhang:2009dd, Nomura:2020zlm, Nomura:2021tmi,Ko:2020qlt}.
With such a mixing, we obtain both a complex vector DM $X^\pm$ and a dark photon $Z'$~\cite{Holdom:1985ag, Fabbrichesi:2020wbt} from the $SU(2)_H$ gauge sector where the stability of DM is guaranteed by remnant $U(1)$ symmetry from custodial one.
In previous works~\cite{Hambye:2008bq, Zhou:2022pom} for a model with minimal field contents, DM-scalar interactions are mainly considered in explaining DM relic density where the new scalar boson plays a role of mediator between the SM and dark sectors via mixing with Higgs boson, but it is also possible to obtain observed relic density via $X^+X^- \to Z' Z'$ process;
Although masses of DM and $Z'$ are degenerate the process is possible in early universe~\cite{DAgnolo:2015ujb}. 
In this case degenerate dark gauge bosons, DM and $Z'$, play a dominant role in dark matter physics and we would get more stringent constraint in gauge sector including kinetic mixing parameter.
Thus it is worth exploring this scenario to constrain gauge coupling, DM/Z' mass and kinetic mixing parameter.

In this work, we discuss $SU(2)_H$ gauge sector introducing only one $SU(2)_H$ doublet scalar and the effective operator providing $SU(2)_H$ and $U(1)_Y$ kinetic mixing.
We then estimate relic density of DM $X^\pm$ focusing on $X^+X^- \to Z' Z'$ process and search for parameter region explaining observation.
In addition, we discuss constraints on the kinetic mixing from dark photon physics and from DM direct detection  by calculating DM-nucleon scattering cross section. 

This paper is organized as follows.
In Sec. II, we introduce the model and formulate mass eigenvalues and eigenstate in scalar/gauge secotor,
and relevant interactions.
In Sec. III, we discuss relic density of DM, constraint from dark photon physics and direct detection of DM.
Finally we provide a conclusion in Sec.IV. 

\section{A model}

We consider a model with hidden sector described by hidden $SU(2)_H$ gauge symmetry where we introduce $SU(2)_H$ doublet scalar field $\Phi$ to break the symmetry spontaneously.
The new scalar field $\Phi$ will develop a vacuum expectation value(VEV) $v_\Phi$ and without the loss of generality it can be written by
\begin{equation}
\Phi = \begin{pmatrix} G^+_H \\ \frac{1}{\sqrt{2}} (v_\Phi + \phi + i G_H)  \end{pmatrix},
\end{equation}
where $G^\pm_H$ and $G_H$ correspond to massless Nambu-Goldstone(NG) boson that are absorbed by $SU(2)_H$ gauge bosons.
In scalar sector we also have the SM Higgs field $H$ written by 
\begin{equation}
H = \begin{pmatrix} w^+ \\ \frac{1}{\sqrt{2}} (v + \tilde h + i G_Z)  \end{pmatrix},
\end{equation}
where $v$ is the VEV of $H$ and $\{w^\pm, G_Z \}$ correspond to NG boson absorbed by the SM massive gauge bosons $\{W^\pm, Z \}$.
Fermion sector is considered to be the same as the SM and we do not focus on fermions in this work.

The renormalizable Lagrangian in the model is 
\begin{equation}
\mathcal{L} = \mathcal{L}_{\rm SM} +(D_\mu \Phi)^\dagger(D^\mu \Phi) -\frac14 X^a_{\mu \nu} X^{a \mu \nu} - \mathcal{V}
\end{equation}
where $X^a_{\mu \nu} (a=1,2,3)$ is the gauge field strength for $SU(2)_H$, $\mathcal{L}_{SM}$ is the Lagrangian of the SM without scalar potential, and $\mathcal{V}$ is the scalar potential including the SM Higgs field $H$.
Explicit form of the scalar potential is given by
\begin{equation}
\mathcal{V} = \mu_H^2 (H^\dagger H) + \mu_\Phi^2  (\Phi^\dagger \Phi) + \lambda_H (H^\dagger H)^2 + \lambda_\Phi (\Phi^\dagger \Phi)^2 + \lambda_{H\Phi} (H^\dagger H)(\Phi^\dagger \Phi).
\label{eq:potential}
\end{equation}
The covariant derivative of $\Phi$ is written by
\begin{equation}
D_\mu \Phi = \left(\partial_\mu + i g_H \frac{\sigma^a}{2} X^a_\mu \right) \Phi
\end{equation}
where $g_H$ is gauge coupling associated with $SU(2)_H$ and $\sigma^a$ is the Pauli matrix acting on representation space of $SU(2)_H$.
In addition to the renormalizable Lagrangian we introduce a gauge invariant effective interaction of
\begin{equation}
\mathcal{L}_{\rm eff} = \frac{1}{\Lambda^2} (\Phi^\dagger \sigma^a \Phi)X^a_{\mu \nu} B^{\mu \nu} 
\label{eq:Leff}
\end{equation}
where $B_{\mu \nu}$ is the gauge field strength of $U(1)_Y$ and $\Lambda$ represents a high energy scale~\footnote{Dimension 8 operator of $(\Phi^\dagger \sigma^a \Phi)X^a_{\mu \nu} (H^\dagger \sigma^b H)W^b_{\mu \nu}$ is also possible that induces mixing between $SU(2)_H$ and $SU(2)_L$~\cite{Zhou:2022pom}. Here we do not consider the operator considering it is more suppressed by $1/\Lambda^4$ factor.}
After $\Phi$ developing its VEV the effective Lagrangian provides kinetic mixing term such that
\begin{align}
\mathcal{L}_{\rm eff} \supset -\frac{v_\Phi^2}{4 \Lambda^2} X^3_{\mu \nu} B^{\mu \nu} & \equiv  -\frac{\epsilon}{2} X^3_{\mu \nu} B^{\mu \nu} \nonumber\\
& = \frac{\epsilon}{2} (\partial_\mu X^3_\nu - \partial_\nu X^3_\mu +  g_H (X^1_\mu X^2_\nu - X^1_\nu X^2_\mu)) B^{\mu \nu},
\end{align} 
indicating mixing between third component of $SU(2)_H$ gauge field $X^3_\mu$ and $U(1)_Y$ gauge field  $B_\mu$.
Kinetic terms of $X^3_\mu$ and $B_\mu$ can be diagonalized by transformation 
\begin{equation}
\left(\begin{array}{c}
X^3_\mu\\
B_\mu\\
\end{array}\right)=\left(\begin{array}{cc}
r & 0 \\
-\epsilon r & 1 \\
\end{array}\right)\left(\begin{array}{c}
\tilde{Z}^\prime_\mu\\
\tilde{B}_\mu\\
\end{array}\right),
\label{eq:kinetic}
\end{equation}
where $r=\frac{1}{\sqrt{1 - \epsilon^2}}$.
We then obtain Lagrangian for gauge sector as 
\begin{equation}
\mathcal{L}_G = - \frac14 X^i_{\mu \nu} X^{i \mu \nu} - \frac14 \tilde{Z}'^{\mu \nu} \tilde{Z}'_{\mu \nu} -\frac14 W^{\alpha \mu \nu} W^\alpha_{\mu \nu} -\frac14 \tilde{B}^{\mu \nu} \tilde{B}_{\mu \nu},
\label{eq:gauge}
\end{equation}
where $i=1,2$ and $W^{\alpha}_{\mu \nu}$ is the gauge field strength of $SU(2)_L$ gauge field $W^{\alpha}_\mu (\alpha =1-3)$. 

\subsection{Scalar sector}

Here we consider scalar sector in the model and derive mass eigenvalues with corresponding eigenstates.
The VEVs  $v$ and $v_\Phi$ can be obtained from the scalar potential Eq.~(\ref{eq:potential}) requiring stationary conditions $\partial V/\partial v = \partial V/\partial v_\Phi = 0$ 
that provide
\begin{align}
& \mu_H^2 +  \lambda_H v^2 + \frac{1}{2} \lambda_{H \Phi} v_\Phi^2 = 0, \\ 
& \mu_\Phi^2 +  \lambda_\Phi v_\Phi^2 + \frac{1}{2} \lambda_{H \Phi } v^2 = 0.  
\end{align}
Here we require $\mu_H^2, \mu_\Phi^2 < 0$ to make square of VEVs positive definite.

After spontaneous symmetry breaking, we obtain mass matrix for CP-even neutral scalar as
\begin{equation}
\mathcal{L} \supset \frac12 \begin{pmatrix} \tilde h \\ \phi \end{pmatrix}^T 
\begin{pmatrix} 2 \lambda_H v^2 & \lambda_{H \Phi} v v_\varphi \\ \lambda_{H \Phi } v v_\Phi & 2 \lambda_\Phi v_\Phi^2 \end{pmatrix}.
\begin{pmatrix} \tilde h \\ \phi \end{pmatrix}
\end{equation} 
We obtain mass eigenvalues by diagonalizing the mass matrix such that
\begin{align}
m_{h,H}^2 = \lambda_H v^2 + \lambda_\Phi v_\Phi^2 \pm \sqrt{(\lambda_H v^2 - \lambda_\Phi v_\varphi^2)^2 + \lambda_{H \Phi }^2 v^2 v_\Phi^2 },
\end{align}
where we identify $m_h = 125$ GeV is the mass of SM Higgs boson.
The mass eigenstates and relevant mixing angle are obtained as 
\begin{align}
& \begin{pmatrix} \tilde h \\ \phi \end{pmatrix} = \begin{pmatrix} \cos \beta & - \sin \beta \\ \sin \beta & \cos \beta \end{pmatrix} \begin{pmatrix} h \\ H \end{pmatrix}, \nonumber \\
& \tan 2\beta = \frac{v v_{\Phi}\lambda_{H \Phi}}{v^2 \lambda_H - v^2_{\Phi}\lambda_\Phi},
\end{align}
where $h$ is identified as the SM Higgs boson.
In the analysis of this paper we assume scalar mixing is negligibly small choosing tiny $\lambda_{H \Phi}$ and focus on the effect of gauge interactions.

\subsection{Gauge boson masses and interactions}
After spontaneous breaking of electroweak and hidden gauge symmetries, we obtain mass terms of electrically neutral gauge fields including the SM ones as follows
\begin{align}
\mathcal{L}_{M} = m_X^2 X^{+ \mu} X^-_{\mu} + \frac12 m^2_{\tilde{Z}' } \tilde{Z}'^{\mu} \tilde{Z}'_{\mu} 
+ \frac12 m^2_{Z_{\rm SM}} \tilde{Z}^\mu \tilde{Z}_\mu + \Delta M^2 \tilde{Z}^\mu \tilde{Z}'_\mu,
\end{align}
where $X_\mu^\pm = \frac{1}{\sqrt{2}} (X^1_\mu \mp i X^2_\mu)$ and $\tilde{Z}_\mu = \cos \theta_W W^3_\mu - \sin \theta_W \tilde{B}_\mu$ with $\theta_W$ being the Weinberg angle.
Squared mass parameters are given by
\begin{align}
& m^2_{Z_{\rm SM}} = \frac{g_Z^2 v^2}{4}, \quad m_X^2 = \frac{g_H^2 v_\Phi^2}{4}, \quad  m_{\tilde{Z}'}^2 = r^2 (m_X^2 +\epsilon^2 m^2_{Z_{\rm SM}} \sin^2 \theta_W  ), \nonumber \\
&  \Delta M^2 = \epsilon r \sin \theta_W m^2_{Z_{\rm SM}}.
\end{align} 
Thus $\tilde{Z}_\mu$ and $\tilde{Z}'_\mu$ fields mix as a consequence of kinetic mixing effect while $X^\pm_\mu$ is mass eigenstate with mass $m_X$.
We obtain mass eigenvalues by diagonalizing mass matrix for $\tilde{Z}$ and $\tilde{Z}'$ as
\begin{equation}
m^2_{Z, Z'} = \frac{1}{2} (m^2_{Z_{\rm SM}}+ m_{\tilde{Z}'}^2) \pm \frac{1}{2} \sqrt{(m^2_{Z_{\rm SM}} - m_{\tilde{Z}'}^2)^2 + 4 \Delta M^4}.
\end{equation}
The mass eigenstates are given by 
\begin{align}
& \begin{pmatrix} Z \\ Z' \end{pmatrix} = \begin{pmatrix} \cos \chi & \sin \chi \\ - \sin \chi & \cos \chi \end{pmatrix} \begin{pmatrix} \tilde{Z} \\ \tilde{Z}' \end{pmatrix}, \\
& \tan 2 \chi =  \frac{ 2 \epsilon \sin \theta_W  m^2_{Z_{\rm SM}}}{m^2_{Z_{\rm SM}} - m^2_{\tilde{Z}}}.
\end{align}
In our analysis we consider $\chi$ to be tiny by choosing $\epsilon$ sufficiently small as the $Z$-$Z'$ mixing is strongly constrained by the electroweak precision tests. 
We thus take $\tan \chi \simeq \sin \chi \simeq \chi$ henceforth.
In addition, gauge boson masses can be approximated as $m_Z \simeq m_{Z_{\rm SM}}$ and $m_{Z'} \simeq m_X$ for small $\epsilon$.

The gauge interactions among $X^\pm$ and $Z'$ can be obtained from Lagrangian Eq.~\eqref{eq:gauge} as
\begin{align}
\label{eq:Zp-X-int}
\mathcal{L}_{X^\pm Z'} = & \ i g_H \Bigl[ (\partial_\mu X^{+}_\nu - \partial_\nu X^+_\mu) X^{- \mu} Z'^\nu - (\partial_\mu X^-_\nu - \partial_\nu X^-_\mu) X^{+ \mu} Z'^{\nu} \nonumber \\
& \qquad + \frac12 (\partial_\mu Z'_\nu - \partial_\nu Z'_\mu)(X^{+\mu} X^{- \nu} - X^{-\mu} X^{+ \nu}) \Bigr] \nonumber \\
& - g_H^2 [X^+_\mu X^{- \mu} Z'_\nu Z'^\nu - X^+_\mu X^-_\nu Z'^\mu Z'^\nu],
\end{align}
where we adopted approximation of $r \simeq 1$ and $\cos \chi \simeq 1$.
In addition $Z'$ can interact with the SM particles through $Z$-$Z'$ mixing. 
The interactions between $Z'$ and SM fermions $f$ can be written by
\begin{equation}
\mathcal{L}_{Z'ff} = \frac{g}{\cos \theta_W} Z'_\mu \bar{f} \gamma^\mu [-\sin \chi (T_3 - Q_f \sin^2 \theta_W) - \cos \chi \epsilon Y_f \sin \theta_W ]f,
\label{eq:Zp-f-int}
\end{equation}
where $T_3$ is diagonal generator of $SU(2)_L$, $Q_f$ is the electric charge and $Y$ is hypercharge.
In the model $Z'$ decays into SM particles while $X^\pm$ is stable and can be DM candidate.
In addition, we have DM-photon and DM-Z interactions from effective term Eq.~\eqref{eq:Leff} such that
\begin{equation}
\mathcal{L}_{\rm eff} \supset - i  g_H \epsilon  \cos  \theta_W ( \partial^{\mu} A^{\nu} - \partial^{\nu} A^{\mu}) X^+_\mu X^-_\nu - i  g_H \epsilon  \sin  \theta_W ( \partial^{\mu} Z^{\nu} - \partial^{\nu} Z^{\mu}) X^+_\mu X^-_\nu,
\label{eq:DM-photon}
\end{equation}
where $A_\mu$ is photon field and we ignore $Z$-$Z'$ mixing effect since it is negligibly small.
We notice that coupling $g_H$ appears instead of electromagnetic coupling $e$ for DM-photon interaction. 

\section{Dark matter phenomenology}

In this section we consider phenomenology of DM such as its relic density and DM-nucleon scattering cross section for direct detection.

\subsection{Relic density}

Relic density of DM is estimated by solving the Boltzmann equation for number density $n_X$ of $X^\pm$,
\begin{equation}
\dot{n}_X + 3 H n_X = \langle \sigma v \rangle (n_{X_{\rm eq}}^2 - n_X^2), 
\end{equation}
where $\langle \sigma \rangle$ is thermal average of DM annihilation cross section, $H$ is Hubble parameter,  and $n_{X_{eq}}$ is number density of $X^\pm$ in thermal equilibrium.
Here we adopt {\it micrOMEGAs} code~\cite{Belanger:2014vza} to estimate relic density by implementing relevant interactions in the model.
In estimating relic density, the relevant parameters in the model are DM(Z') mass $m_X$, new gauge coupling $g_H$ and scalar mass $m_\phi$. 
In this work we consider $\phi$ is heaver than $X^\pm$ and $Z'$ so that a DM annihilation mode including $\phi$ in final states is forbidden kinematically.
Thus the DM annihilation process in our scenario is solely 
\begin{equation}
X^+ X^- \to Z' Z'.
\end{equation}
Here we fix $m_\phi = 1.5 m_X$ for simplicity and search for parameter region on $\{m_X, g_H \}$ plane that can provide observed DM relic density.
Here we consider two mass range of DM (dark photon) as follows
\begin{align}
& {\rm (i)} \ m_X = [1.0, 60] \ {\rm GeV} \\
& {\rm (ii)} \ m_X =[150, 3000] \ {\rm GeV}. 
 \end{align}
 Here the range (i) is light WIMP like DM case where $Z'$ is also lighter than $Z$ and more dark photon like.
 The range (ii) corresponds to heavy WIMP DM case and $Z'$ is heavier than $Z$.
 We do not consider the region $m_X \sim m_Z$ since $Z$-$Z'$ mixing tends to be large.
 
\begin{figure}[tb]
 \begin{center}
\includegraphics[width=7cm]{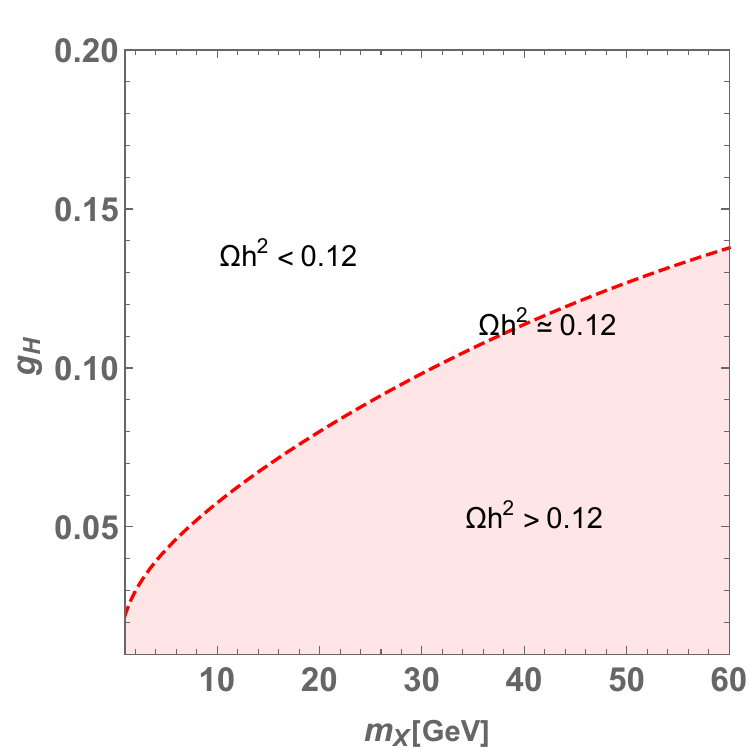} \quad
\includegraphics[width=7cm]{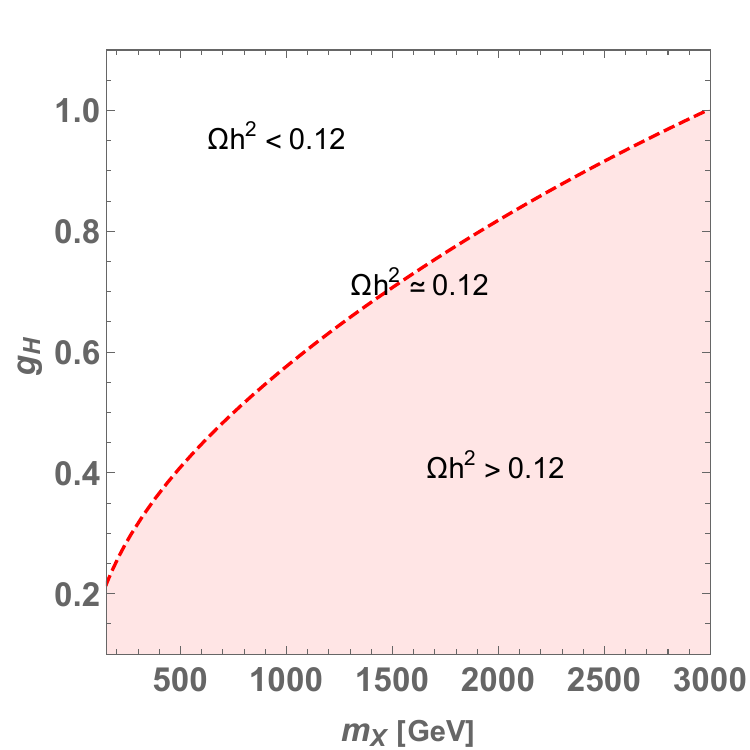} 
\caption{The red dashed curve indicate parameter $\{m_X, g_H \}$ realizing observed relic density $\Omega h^2 \simeq 0.12$. Shaded region is excluded since DM is over abundant. }
\label{fig:relic1}
\end{center}
\end{figure}

 In Fig.~\ref{fig:relic1}, we show the region providing $\Omega h^2 < 0.12$, $\Omega h^2 \simeq 0.12$ and $\Omega h^2 > 0.12$ on $\{m_X, g_H \}$ plane.
 The red dashed curve indicate the parameters explaining observed relic density.
 The shaded region is excluded since relic density is over abundant while the white region provides smaller relic density.
 We find that observed relic density can be obtained by gauge coupling $g_H \in [0.01, 1]$ for DM mass of 1 GeV to 3000 GeV.
 
\subsection{Constraints on kinetic mixing}

Here we summarize constraints on kinetic mixing parameter $\epsilon$ from dark photon searches, thermalization condition of DM and mili-charged DM interaction. \\
\noindent
{\it Constraints from dark photon search}: \\
In the mass range (i) of our interest $\epsilon$ is constrained by BaBar~\cite{BaBar:2014zli, BaBar:2017tiz} and LHCb~\cite{LHCb:2019vmc} experiments that search for dark photon.
The BarBar experiment search for the process $e^+ e^- \to \gamma Z'$ where $Z'$ is considered to decay into visible modes $\{e^+ e^-, \mu^+ \mu^-, \text{light mesons} \}$ or 
invisible mode such as neutrinos and DM states.
The LHCb experiment search for the dark photon which is produced in proton-proton collision and decays into $\mu^+ \mu^-$ pair.
We apply constraint from these experiments for the allowed range of $\epsilon$ where we use the {\it DARKCAST} code to extract the limit~\cite{Ilten:2018crw}.

\noindent
{\it DM thermalization condition}: \\
We require particles in dark sector, DM and dark photon, are in thermal equilibrium before freeze-out of DM to guarantee our relic density calculation.
In addition, dark photon lifetime should be smaller than $\mathcal{O}(1)$s to avoid inconsistency with Big Bang Nucleosynthesis.
In the model, dark sector is thermalized in the early Universe via dark photon decays into electrons.  
Then we impose the condition where the decay rate is larger than the Hubble rate at freeze-out.
This condition can be written by~\cite{Pospelov:2007mp, delaVega:2023dmw}
\begin{equation}
|\epsilon| \gtrsim 2.0 \times 10^{-8} \sqrt{\frac{ 10 \ {\rm GeV} }{m_{Z'}}  } \left( \frac{m_X}{10 \ {\rm GeV}} \right).
\end{equation} 
We adopt the condition as a lower bound of $\epsilon$ in the analysis below.

\noindent
{\it Constraint from mili-charged DM interaction}: \\
Unlike to the Abelian hidden $U(1)$ case we have DM-photon interaction Eq.~\eqref{eq:DM-photon} which arise from the effective interaction Eq.~\eqref{eq:Leff}.
We thus comment on constraint from DM-Baryon scattering for milli-charged DM~\cite{Xu:2018efh} through the interaction.
In this case the cross section is $\sigma_X = \sigma_{X0} v^2$ where $v$ is DM-Baryon relative velocity.
It is suppressed by $v^2$ factor compared to milli-charged fermion DM case $(\sigma_F = \sigma_{F0} v^4)$ that is due to 
momentum factor appearing from derivative in the interaction Eq.~\eqref{eq:Leff}.
Therefore the constraint is not severe and DM direct detection experiment provide stronger constraint that is estimated below.


\subsection{Direct detection}

In the model DM $X^\pm$ interacts with nucleon via $Z'$ exchanging process through interactions in Eqs.~\eqref{eq:Zp-X-int} and \eqref{eq:Zp-f-int}. 
In addition $X^\pm$ have interaction with the SM $Z$ boson via $Z$-$Z'$ mixing where relevant interaction for DM-nucleon scattering is 
\begin{align}
\mathcal{L}_{Z X^+ X^-} = & \ i g_H \sin \chi \Bigl[ (\partial_\mu X^{+}_\nu - \partial_\nu X^+_\mu) X^{- \mu} Z^\nu - (\partial_\mu X^-_\nu - \partial_\nu X^-_\mu) X^{+ \mu} Z^{\nu} \nonumber \\
& \qquad + \frac12 \left( 1- \frac{\epsilon}{\sin \chi} \right) (\partial_\mu Z_\nu - \partial_\nu Z_\mu)(X^{+\mu} X^{- \nu} - X^{-\mu} X^{+ \nu}) \Bigr], 
\end{align} 
where we included Eq.~\eqref{eq:DM-photon} for the third term inside the bracket.
Thus we need to take into account both $Z'$ and $Z$ boson exchanging diagrams to obtain DM-nucleon scattering cross section.
Here vector type interaction among $Z(Z')$ and nucleon can be obtained as 
\begin{align}
& \mathcal{L}_{N} = \sum_{N=p,n} \frac{g}{\cos \theta_W} \bar{N}  (C_{ZNN}^V Z_\mu + C_{Z'NN}^V Z'_\mu ) \gamma^\mu N, \\
& C_{Zpp}^V =  \frac{1}{4} - \sin^2 \theta_W, \quad C_{Znn}^V =   -\frac14, \\
& C_{Z'pp}^V = -\sin \chi \left( \frac{1}{4} - \sin^2 \theta_W \right) - \frac13 \epsilon \sin \theta_W, \quad C_{Z'nn}^V =  \frac14 \sin \chi - \frac12 \epsilon \sin \theta_W,
\end{align} 
where $p$ and $n$ respectively indicate proton and neutron, and we approximate $\cos \chi \simeq 1$. 
We then calculate diagrams and obtain DM-nucleon scattering cross section in non-relativistic limit such that
\begin{equation}
\sigma_{NX} = \frac{2 g_H^2 g^2}{\pi \cos^2 \theta_W} \left( \frac{m_X m_N}{m_X + m_N} \right)^2 \left( \frac{\sin \chi C^V_{ZNN}}{m_Z^2} + \frac{\cos \chi C^V_{Z'NN}}{m_{Z'}^2} \right)^2,
\end{equation}
where $m_N$ denote nucleon mass $(N=p, n)$.
Here we consider average of DM-nucleon scattering cross section $(\sigma_{nX}+\sigma_{pX})/2$ and compare it with experimental constraints 
to estimate upper limit of $\epsilon$. 
In estimating the cross section we adopt $g_H$ value explaining relic density $\Omega h^2 \simeq 0.12$ in Fig.~\ref{fig:relic1} for each DM (dark photon) mass.

 \begin{figure}[tb]
 \begin{center}
\includegraphics[width=7cm]{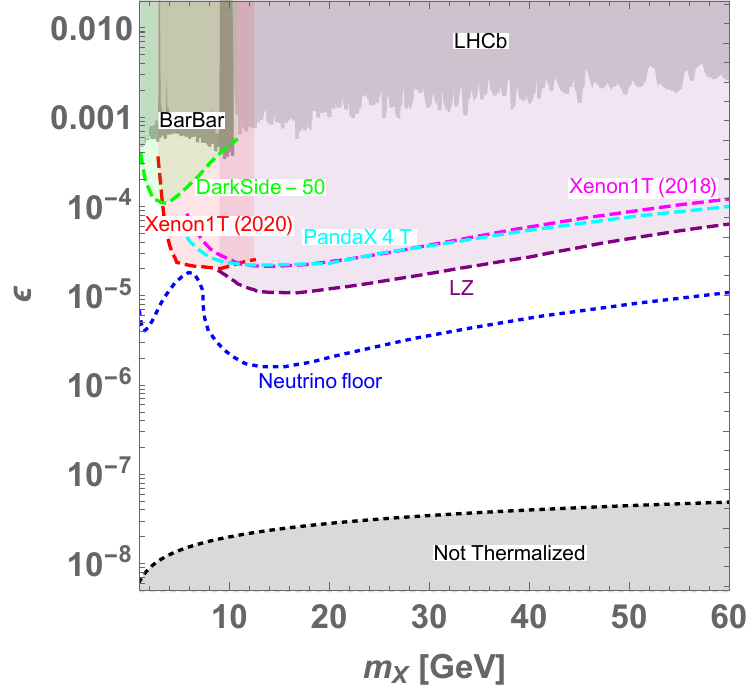} \quad 
\includegraphics[width=7cm]{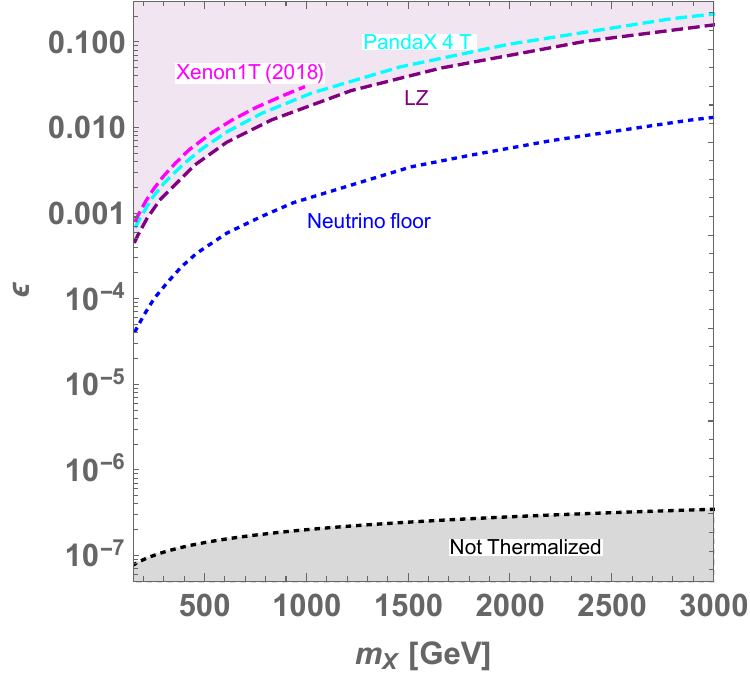} 
\caption{The constraints on $\epsilon$. The colored region above curves are excluded by direct detection experiments indicated by labels. 
In the gray region labeled as "Not Thermalized", $\epsilon$ is too small so that DM and dark photon are not thermalized before freeze out. 
On the left plot, region labeled as LHCb and BarBar are excluded by dark photon searches by these experiments. 
The $\epsilon$ value below blue dotted curve provides DM-nucleon cross section in neutrino floor.}
\label{fig:epsilon}
\end{center}
\end{figure}

In Fig.~\ref{fig:epsilon}, we show constraints on $\epsilon$ on $\{m_X, \epsilon \}$ plane. 
The colored region above curves are excluded by direct detection experiments indicated by labels where the strongest limits are from LZ~\cite{LZ:2022lsv}, Xenon1T (2020)~\cite{XENON:2019gfn} and DarkSide-50~\cite{DarkSide-50:2022qzh} for mass range $m_X \gtrsim 10$ GeV,  3 GeV $\lesssim m_X \lesssim$ 10 GeV and 1 GeV $\lesssim m_X \lesssim$ 3 GeV respectively; we also show constraints from Xenon1T(2018)~\cite{XENON:2018voc} and PandaX-4T~\cite{PandaX-4T:2021bab}. In the gray region, $\epsilon$ is too small so that DM and dark photon are not thermalized before freeze out. The $\epsilon$ value below blue dotted curve provides DM-nucleon cross section in neutrino floor, and it is difficult to explore by direct detection. 
In the left plot, we also indicate excluded region by dark photon searches by LHCb and BarBar.
We find that the direct detection cross section gives us much stronger constraints than dark photon searches when DM (dark photon) is relatively light. 
Future DM direct detection experiments can also explore the allowed region unless it is above the region of neutrino floor.

\section{Summary}

We have discussed DM and dark photon from a hidden $SU(2)_H$ model with minimal new field content which is one $SU(2)_H$ doublet scalar fields, 
introducing dimension 6 effective operator among $SU(2)_H$ and $U(1)_Y$ gauge fields and the doublet.
The effective operator induces kinetic mixing between these gauge fields after spontaneous gauge symmetry breaking and one component of $SU(2)_H$ gauge field becomes
dark photon $Z'$ while the other two components become vector DM $X^\pm$.
Note that $X^\pm$ and $Z'$ have almost degenerate mass in the model.

Then we have considered the scenario in which relic density of DM is explained by $X^+ X^- \to Z' Z'$ process since it is determined by gauge interaction only and more constrained due to fixed mass relation between $X^\pm$ and $Z'$. 
Numerically analyzing the Boltzmann equation, we searched for parameter region that satisfy observed relic density.
It has been found that $\mathcal{O}(0.01)$ to $\mathcal{O}(1)$ gauge coupling of $SU(2)_H$ can realize the relic density in the DM (dark photon) mass range of $\sim$1 GeV to $\sim 3$ TeV.
We have also discussed relevant constraints from DM and dark photon physics to constrain the kinetic mixing parameter.
In addition, cross section of DM-nucleon scattering via dark photon exchange is estimated to explore constraints from DM direct detection searches.
As a result, we have found the direct detection searches can provide the strongest constraint on kinetic mixing parameter and 
further parameter space will be tested in future unless the cross section is above neutrino floor.

\section*{Acknowledgments}
The work was also supported by the Fundamental Research Funds for the Central Universities (T.~N.).

\end{document}